\documentclass[twocolumn]{aastex62}
\graphicspath{{./}}

\submitjournal{ApJ}

\shorttitle{Sample article}
\shortauthors{Li et al.}

\begin{document}
% Trigger dense molecular core collapsing by core-core collision mechanism: The case of Barnard 68
\title{Evidence for Core-Core Collision in Barnard 68}

\correspondingauthor{Dalei Li}
\email{lidalei@xao.ac.cn}

\author[0000-0002-0786-7307]{Dalei Li}
\affil{Xinjiang Astronomical Observatory, Chinese Academy of Sciences, Urumqi 830011, People's Republic of  China}
\affil{University of Chinese Academy of Sciences, Beijing 100049, People's Republic of  China}
%\affil{Key Laboratory of Radio Astronomy, Chinese Academy of Sciences, Urumqi 830011, People's Republic of  China}
\affil{Xinjiang Key Laboratory of Radio Astrophysics, Urumqi 830011, People's Republic of  China}
\author[0000-0002-1363-5457]{Christian Henkel}
\affil{Max-Planck-Institut f$\ddot{u}$r Radioastronomie, Auf dem H$\ddot{u}$gel 69, D-53121 Bonn, Germany}
\affil{Xinjiang Astronomical Observatory, Chinese Academy of Sciences, Urumqi 830011, People's Republic of  China}
\author[0000-0002-1363-5457]{Alexander Kraus}
\affil{Max-Planck-Institut f$\ddot{u}$r Radioastronomie, Auf dem H$\ddot{u}$gel 69, D-53121 Bonn, Germany}
\author[0000-0002-0786-7307]{Xindi Tang}
\affil{Xinjiang Astronomical Observatory, Chinese Academy of Sciences, Urumqi 830011, People's Republic of  China}
\affil{University of Chinese Academy of Sciences, Beijing 100049, People's Republic of  China}
\affil{Xinjiang Key Laboratory of Radio Astrophysics, Urumqi 830011, People's Republic of  China}
\affil{Key Laboratory of Radio Astronomy and Technology,Chinese Academy of Sciences, A20 Datun Road, Chaoyang District, Beijing, 100101, P. R. China}
\author[0000-0002-0786-7307]{Willem Baan}
\affil{Xinjiang Astronomical Observatory, Chinese Academy of Sciences, Urumqi 830011, People's Republic of  China}
\author[0000-0002-0786-7307]{Jarken Esimbek}
\affil{Xinjiang Astronomical Observatory, Chinese Academy of Sciences, Urumqi 830011, People's Republic of  China}
\affil{University of Chinese Academy of Sciences, Beijing 100049, People's Republic of  China}
\affil{Xinjiang Key Laboratory of Radio Astrophysics, Urumqi 830011, People's Republic of  China}
\affil{Key Laboratory of Radio Astronomy and Technology,Chinese Academy of Sciences, A20 Datun Road, Chaoyang District, Beijing, 100101, P. R. China}
\author[0000-0002-0786-7307]{Ke Wang}
\affil{Kavli Institute for Astronomy and Astrophysics, Peking University, Beijing 100871, People's Republic of  China}
\author[0000-0002-0786-7307]{Gang Wu}
\affil{Xinjiang Astronomical Observatory, Chinese Academy of Sciences, Urumqi 830011, People's Republic of  China}
%\affil{Key Laboratory of Radio Astronomy, Chinese Academy of Sciences, Urumqi 830011, People's Republic of  China}
\affil{Xinjiang Key Laboratory of Radio Astrophysics, Urumqi 830011, People's Republic of  China}
\author[0000-0002-0786-7307]{Tie Liu}
\affil{Shanghai Astronomical Observatory, Chinese Academy of Sciences, 80 Nandan Road, Shanghai 200030, People's Republic of  China}
\author[0000-0002-0786-7307]{Andrej M. Sobolev}
\affil{Ural Federal University, 19 Mira Street, 620002 Ekaterinburg, Russia}
\affil{Xinjiang Astronomical Observatory, Chinese Academy of Sciences, Urumqi 830011, People's Republic of  China}
\author[0000-0002-0786-7307]{Jianjun Zhou}
\affil{Xinjiang Astronomical Observatory, Chinese Academy of Sciences, Urumqi 830011, People's Republic of  China}
%\affil{University of Chinese Academy of Sciences, Beijing 100049, People's Republic of  China}
\affil{Xinjiang Key Laboratory of Radio Astrophysics, Urumqi 830011, People's Republic of  China}
\affil{Key Laboratory of Radio Astronomy and Technology,Chinese Academy of Sciences, A20 Datun Road, Chaoyang District, Beijing, 100101, P. R. China}
\author[0000-0002-0786-7307]{Yuxin He}
\affil{Xinjiang Astronomical Observatory, Chinese Academy of Sciences, Urumqi 830011, People's Republic of  China}
\affil{University of Chinese Academy of Sciences, Beijing 100049, People's Republic of  China}
%\affil{Key Laboratory of Radio Astronomy, Chinese Academy of Sciences, Urumqi 830011, People's Republic of  China}
\affil{Xinjiang Key Laboratory of Radio Astrophysics, Urumqi 830011, People's Republic of  China}
\author[0000-0002-0786-7307]{Toktarkhan Komesh}
\affil{Energetic Cosmos Laboratory, Nazarbayev University, Astana 010000, Kazakhstan}
\affil{Institute of Experimental and Theoretical Physics, Al-Farabi Kazakh National University, Almaty 050040, Kazakhstan}

\begin{abstract}
The prestellar core Barnard 68 (B68) is a prototypical source to study the initial conditions and chemical processes of star formation. A previous numerical simulation suggested the southeastern bullet is impacting on the 
main body of B68. In order to obtain more observational evidence, mapping observations of 
the ground state SO\,($1_0-0_1$) emission line  at 30\,GHz 
were made with the Effelsberg 100\,m  telescope. 
Based on the velocity field and channel maps derived from SO, 
three velocity components were clearly detected. 
%The southeastern  bullet presents the most redshifted %and the main body 
%shows blueshifted emission in the east and more %redshifted one in the west. 
The velocity field of the main body indicates rotation and is well fitted by a solid-body rotation model. 
The measured radial velocity difference between the bullet and the main core is about 
0.4\,km\,s$^{-1}$,  which is almost equal to the velocity obtained by the previous numerical simulation. 
Therefore, the bullet is most likely  impacting onto the rotating main body of B68.  % smaller core
A 1D spherical non-LTE Monte-Carlo radiation transfer RATRAN code is performed to derive the radial abundance profile of SO by analyzing the observed velocity-integrated intensity. 
%Using non-LTE radiation transfer RATRAN code, the radial  abundance profile of SO is derived. 
SO is depleted inside a 60$^{\prime\prime}$ (0.02\,pc) radius from the core. 
The abundance stays constant at 2.0$\times$10$^{-9}$ for radii larger than 60$^{\prime\prime}$ from the center of the main core. 
The abundance is enhanced at the interface of the bullet and the main core indicating that shock waves were produced by the collision between the bullet and the main core. 
In conclusion, based on the kinematical and chemical analysis, our observational results support the previously proposed core-core collision scenario in B68.

% From a simulation, 
% We consider that the collision can produce shock waves heating the gas and leading to increased central gas temperatures. 
\end{abstract}

\section{Introduction}

Stars are generally believed to be formed from dense collapsing  molecular cores. Observations towards dense molecular cores can help us to understand the initial conditions and early stages of star formation \citep{Bergin2007, MeKee2007}.  
Especially, isolated dense molecular cores are ideal targets for testing star-forming theories, such as Barnard 68 \citep[B68;][]{Barnard1919}, which is a small and isolated Bok globule \citep{Bok1947}. Because of  its simple and at first sight almost spherical structure, B68 is a perfect source to study the initial conditions of star formation.
% Performing near-infrared observations towards B68 and based on dust extinction method,

Based on near-infrared observations and dust extinction studies \citep{Lada1994},  
\citet{Alves2001} derived the column density structure of B68 with high sensitivity and spatial resolution. 
%(5$^{\prime\prime}$). 
They then found that the  column density structure is well fitted by a Bonnor-Ebert sphere \citep{Bonnor1956, Ebert1955} with a fitted parameter $\xi_{\rm max}$\,=\,6.9\,$\pm$\,0.2 
\citep[][see Appendix A]{Alves2001}. This model involving thermal pressure dominating  against gravity  was supported by follow-up molecular line observations. 
For example, NH$_{3}$ data observed with the Effelsberg 100\,m telescope \citep{Hotzel2002} show extremely narrow full width to half maximum (FWHM) line widths of 0.181\,$\pm$\,0.003\,km\,s$^{-1}$.
%in which the non-thermal component accounts for less than 10\% considering a gas kinetic temperature of 10\,K \citep{Hotzel2002,Lai2003}. 
Such narrow line widths were also detected by follow-up observations of other molecules in B68, for 
 C$^{18}$O, about 0.18\,$\pm$\,0.01\,km\,s$^{-1}$, and for C$^{34}$S, about 0.15\,$\pm$\,0.01\,km\,s$^{-1}$ \citep{Lada2003}. The estimated sonic Mach numbers (ratios of the nonthermal velocity disperson with respect to sound speed),  are 
 0.17 for NH$_{3}$, 0.30 for C$^{18}$O and 0.25 for C$^{34}$S assuming the gas kinetic temperature of 10\,K.
 Distance estimates for B68 range from 60 to 205\,pc \citep{deGeus1989, Hotzel2002, Lai2003, Lombardi2006, Alves2007}. In order to fit the B68 density structure with a Bonnor-Ebert sphere at 10\,K, a distance of 78\,pc is adopted in the following \citep{Hotzel2002, Lai2003, Li2018}.
% indicating that B68 is mainly supported by thermal pressure. the kinetic gas temperature of B68 was determined to be 10 -\,- 11\,K measured by NH$_3$ observations \citep{2002A&A...395L...5H, 2003AJ....126..311L}.

%  However, the fitted parameter $\xi_{\rm max}$ = 6.9$\pm$0.2 is a little bit larger than the critical value of $\xi_{\rm max}$ = 6.5,  indicating that B68 may be unstable \citep{Li+18}.

The internal motions and  dynamical state of B68 can be deduced from molecular line observations.
%emission line observations,
\citet{Lada2003} performed high spatial resolution C$^{18}$O (1-0), N${_2}$H$^{+}$(1-0), C$^{32}$S (2-1) and  C$^{34}$S (2-1)  molecular emission line mapping observations using the IRAM (Institut de Radioastronomie Millim{\'e}trique) 30\,m telescope. In contrast to the C$^{34}$S single peak line profiles, C$^{32}$S (2-1) shows complex blueshifted and red-shifted profiles  so-called 'self-reversed' line structure revealing that B68 contains very complex motions indicating non-radial oscillations \citep{Lada2003}.
%Nonradial oscillation was suggested. 
Soon after, the non-radial oscillations were also  detected by follow-up  JCMT (James Clerk Maxwell Telescope) HCO$^{+}$(3-2) mapping observations \citep{Redman2006}. 
In additional studies, many molecular lines were detected including  $^{13}$CO, C$^{18}$O, HCO$^{+}$, H$^{13}$CO$^{+}$, DCO$^{+}$, N${_2}$H$^{+}$ and more complex motions inside the core were revealed \citep{Maret2007}. For example, at the core center, they found that molecular transitions tracing particularly dense gas, HCO$^{+}$ (4-3), DCO$^{+}$ (3-2) and N${_2}$H$^{+}$ (1-0), show a red asymmetry indicating expansion while  HCO$^{+}$ (1-0) and (3-2) show a blue asymmetry,  revealing  radial oscillations at the core center of B68 \citep{Maret2007}.
%If the complex spectral line profiles were interpreted as 
Because B68 is experiencing non-radial and radial oscillations  \citep{Keto2006,Redman2006,Maret2007}, its core could be more stable and much more long-lived, about 3$\times$10$^{6}$ years, than suggested by the free-fall timescale of 1.7$\times$10$^{5}$ years \citep{Burkert2009}. 

However, as mentioned above, the fitted parameter $\xi_{\rm max}$\,=\,6.9\,$\pm$\,0.2 \citep{Alves2001} is slightly larger than the critical value of $\xi_{\rm max}$\,=\,6.5 for a Bonnor-Ebert sphere model indicating that B68 may be unstable and collapse with a freefall timescale \citep{Burkert2009,Li2018}. In order to resolve this contradiction, \citet{Burkert2009} proposed a core-core collision scenario and  
did a simulation that the smaller core at the southeastern corner of B68  is impacting on the  main body, initiating and accelerating gravitational collapse. In the numerical simulation,  the impact velocity is with 0.37\,km\,s$^{-1}$ supersonic and the supersonic fluid may produce shock waves.
%For testing the core-core collision scenario, we proposed a shock tracer SO molecular line mapping observations towards the %entire region of B68. 

SO  is an  excellent tracer to study a cold dark cloud and to trace shock waves. There are many cold dark clouds already mapped by the ground transition SO\,($1_0-0_1$). For example, \citet{Rydbeck1980} performed SO observations toward many dark molecular clouds with a 2$^{\prime}$ resolution, \citet{Heithausen1995} inspected galactic cirrus, \citet{Codella1997} studied Bok globules and \citet{Lique2006} observed TMC-1. Besides, SO is also a good species to trace shock waves which can significantly enhance the SO abundance.
%(Schmid-Burgk et al.1992; Pineau des For$\hat{e}$ts et al. 1993).
By performing SO observations, \citet{Martin1992} already mapped L1448 and Cep A in both ambient quiescent and shocked outflow regions and found [SO/H$_2$] ratios being enhanced, with fractional abundances of quiescent gas,  $0.9\times 10^{-9}$, to that of shocked regions, $>15\times 10^{-9}$ for L1448 and $1.8\times 10^{-9}$ to $>13\times 10^{-9}$ for Cep A \citep{Blake1987,Schilke1997, Comito2005}.
\citet{Friberg1984} mapped the Orion KL region using the SO emission and found that the SO abundance is enhanced to 2$\times$ 10$^{-6}$ in the active regions of Orion KL, while the ambient abundance of SO is only 0.3\% that of the active regions.
%6$\times$ 10$^{-9}$ 
%in the ambient quiescent region. 
\citet{Schmid1992} even found strong SO emission where CO is absent, located in some regions of the blue lobe of the L1221 outflow. There exists a strong correlation between SO spots and the endpoints of molecular outflows \citep{Schmid1994,Chuang2021}. 
Therefore, 
SO is likely a potential ideal tracer to 
test the core-core collision scenario
towards the prototypical Bok globule B68.
% Lee2023
% In conclusion, SO is therefore an ideal tracer to study our prototypical Bok globule B68
% % if it really undergoes 
% to reveal the dynamic collision.

\section{Observations and data archive}
Mapping observations of the SO\,(1$_{0}$\,-\,0$_{1}$) molecular emission line  at 30.00154\,GHz\footnote{
This value was taken from the Spectral Line 
Atlas of Interstellar Molecules (SLAIM) (Available at 
https://splatalogue.online//$\#$/basic).}(F.J.Lovas, private communication, Remijan et al. 2007)
%\citep{Gottlieb1978} 
were made with the  Effelsberg 100\,m telescope\footnote{This work is based on observations with the 100-m telescope of the MPIfR (Max-Planck-Institut f$\ddot{\rm u}$r Radioastronomie) at Effelsberg.} by adopting  the On-The-Fly (OTF)  in position switching mode in March, 2017. The spatial angular resolution is 30$^{\prime\prime}$. In order to increase the signal-to-noise ratios in individual channels, the data were smoothed to 40$^{\prime\prime}$ ($\sim$\,0.015\,pc).
The prime focus receiver P10mm (4-Box) with a frequency coverage of 
27.0\,$--$\,38.5\,GHz
%27.0\,-\,-\,38.5\,GHz 
was selected.
An XFFT (FFT: Fast Fourier Transform) spectrometer 
with 32768 spectral channels and a bandwidth of 400\,MHz, corresponding to a channel width of 0.036\,km\,s$^{-1}$, has been employed. The 
data were reduced by the CLASS package of GILDAS\footnote{http://www.iram.fr/IRAMFR/GILDAS}.
All spectra were individually corrected for the atmospheric attenuation and the gain-elevation effect of the antenna (loss of sensitivity due to gravitational deformation during tilt).
For the final calibration into K, $T_{\rm MB}$ conversion factors were derived by observing suitable calibrators (NGC\,7027) in continuum \citep{Winkel2012}.
The uncertainties are of the order of 10\%.
%5\,-\,10\%, depending 
%on weather conditions.

In addition to the molecular SO line data, a DSS (Digitized Sky Surveys) image was retrieved from the ESO data archive\footnote{https://archive.eso.org/dss/dss}.

%The higher transition of SO\,(3$_{2}$\,-\,2$_{1}$, 99.2999\,GHz) %emission line mapping observations were  carried out  with the Delingha %13.7\,m telescope located in Qinghai province, China. 
%The spatial angular resolution is about 60$^{\prime\prime}$.
%The bandwidth of the backend is 200\,MHz, resulting in a velocity %resolution  0.037\,km\,s$^{-1}$. 
%The main beam efficiency is 0.62. The data is converted to $T_{\rm mb}$.

\section{Results}
% Fig. 1
% spectra line map -> no self-reversed line profiles 
% Fig. 2
% moment maps -> velocity-integrated intensity map, 
% velocity field 
% line width 
% Fig. 3
% channel maps -> 
% Fig. 4
% position velocity diagram ->
% rotation fit -> velocity field - fitted results -> residual 
% Fig. 5 
% run ratran to derive the abundance 
% estimate along several directions to plot intensity 
% two lines to support core-core collision 
% dynamics + chemistry 
\subsection{The spectral SO lines}
\begin{figure}[t]
% \vspace*{0.2mm}
\begin{center}
\includegraphics[width=130mm, trim={0 0 0 0},clip]{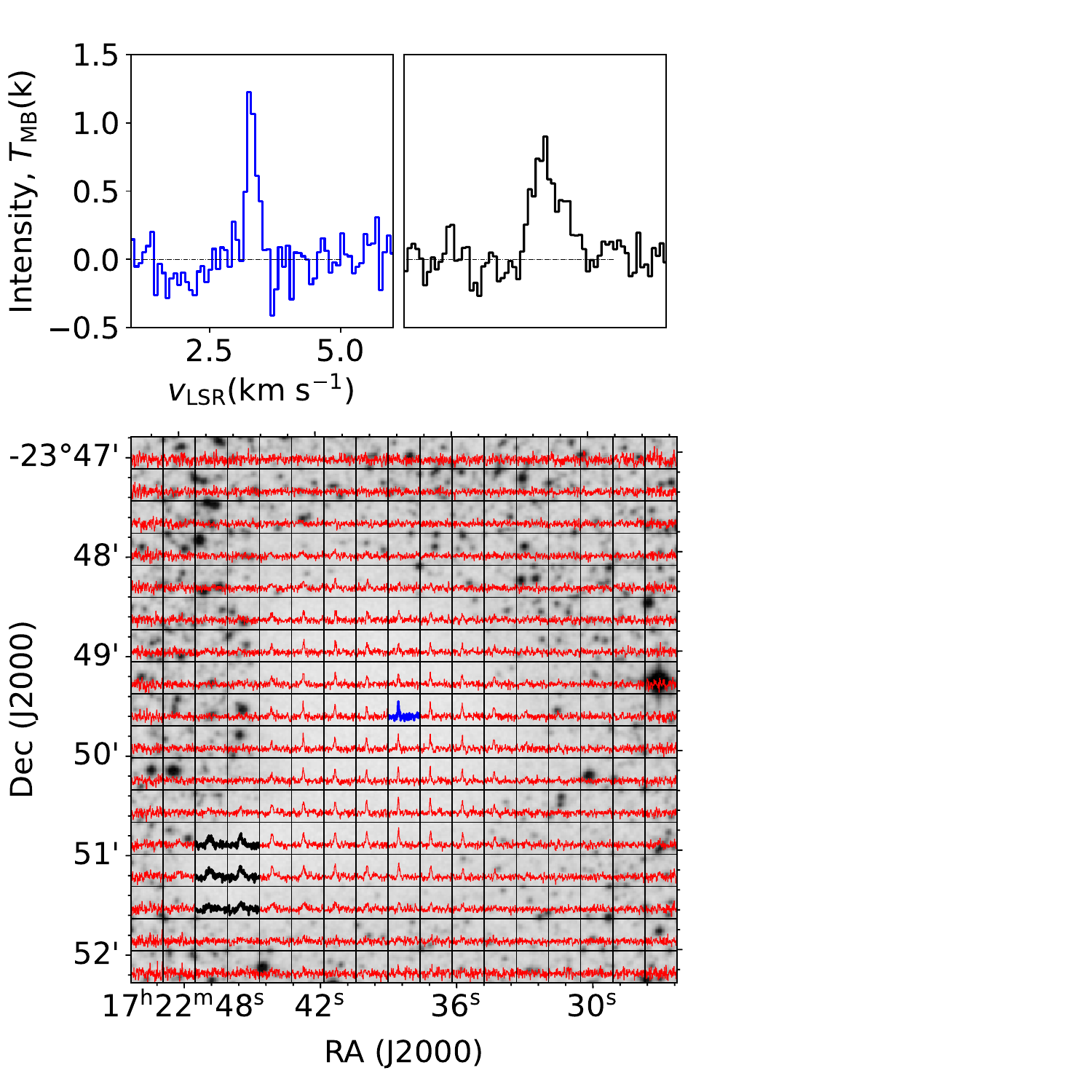}%,angle=-90
\end{center}
\caption{Observed SO spectra. 
Upper panel (left): The SO spectrum towards the column density peak ($\alpha_{\rm J2000}$\,=\,17:22:38.60, $\delta_{\rm J2000}$\,=\,-23:49:46.0)
% \citep[
% $\alpha_{\rm J2000}$\,=\,17:22:38.60, $\delta_{\rm J2000}$\,=\,-23:49:46.0;
% ]
% []{Bergin2002}
presented with a Local Standard of Rest velocity scale. 
Upper panel (right): The averaged spectra derived from six spectral lines (black) in the tail region of the core. The x- and y-axis scales are identical to those of the spectra on the left.
Bottom panel: The spectral line map overlaid on a negative Digital Sky Survey  (DSS) image, with white indicating the morphology of the globule and black representing stellar objects in the outskirts of the image. 
The profiles of the upper panel are emphasized in the middle and at the bottom left of the lower panel, respectively.
}
\label{spectra_maps}
\end{figure}

The upper panel in Fig.\,\ref{spectra_maps} presents the observed spectral line of SO towards the dust extinction peak of B68 \citep{Bergin2002}. 
Contrary to the complex line profiles detected in
HCO$^+$\,(1\,-\,0), (3\,-\,2), (4\,-\,3), 
DCO$^+$\,(2\,-\,1) and (3\,-\,2) \citep[see Fig\,.1 of ][]{Maret2007}, 
the SO shows a single profile. 
% trim={<left> <lower> <right> <upper>}

The lower panel in  Fig.\,\ref{spectra_maps} shows the observed SO line profiles. Both the main body and the southeastern bullet are clearly detected. Not only the spectrum towards the column density peak, but all spectra show single Gaussian profiles, which are quite different from the previous mapping 
results of  asymmetric, self-reversed line profiles as seen in the CS\,(2\,-1\,) and HCO$^+$\,(3\,-\,2) lines obtained by the IRAM 30\,m and JCMT 15\,m, respectively \citep{Lada2003, Redman2006}.

% red and blue asymmetric line profiles 
% Compared to previous molecular line mapping results, exhibit
\subsection{The spatial distribution of SO}
\begin{figure*}[t]
% \vspace*{0.2mm}
\begin{center}
\includegraphics[width=180mm]{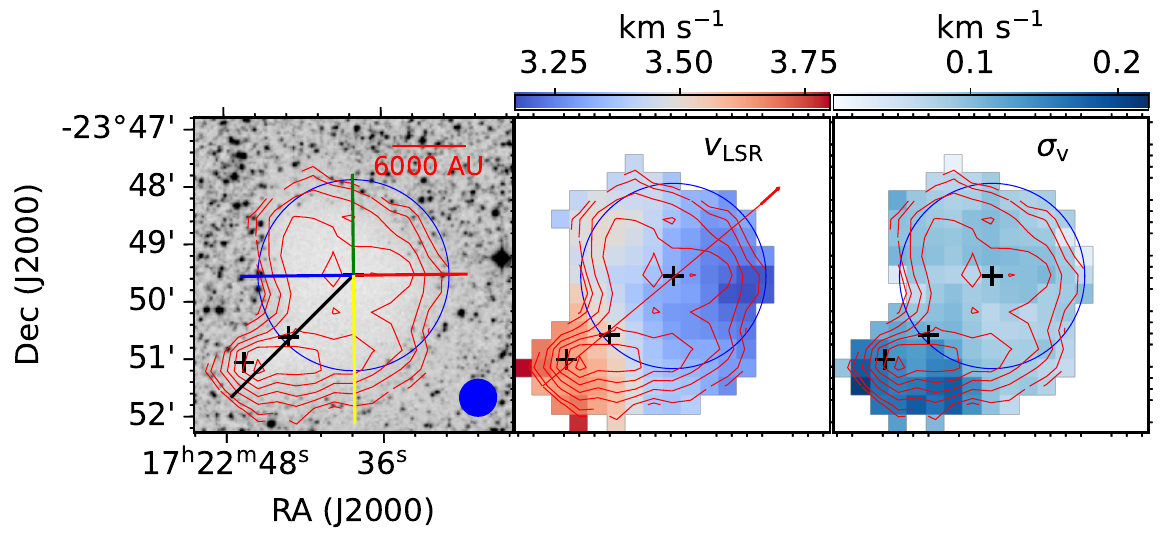}%,angle=-90
\end{center}
\caption{The spatial distribution of SO. Velocity-integrated intensity contour map of B68 overlaid on the Digitized Sky Survey (DSS) image. 
The blue circle with a radius of 100$
^{\prime\prime}$ marks the main body of B68 \citep{Alves2001}.
The plus symbols show the peaks of the three dust cores detected by the dust extinction measurements with an angular resolution of 10$^{\prime\prime}$ \citep[see Fig\,.7 dust extinction map of ][]{Alves2001M}.
Left: The red contours show the SO(1$_{0}$ - 0$_{1}$) molecular line emission. 
The contour levels start from 2$\sigma$ ($\sigma$\,=\,0.036\,K\,km\,s$^{-1}$) with an interval 2$\sigma$. 
The beam size (blue filled circle) is shown in the lower right corner of the left panel. The various straight lines show the paths to extract the velocity-integrated intensity distribution presented in Fig.\,\ref{abundance}. The middle panel marks the velocity field distribution. The red line is the path for the position-velocity diagram in Fig.\,\ref{pv_diagram}. The right panel presents the velocity dispersion $\sigma_{\rm v}$. 
} 
%$\sigma$ ($\sigma$ = K km/s).
\label{int_maps}
\end{figure*}

\begin{figure}[t]
\vspace*{0.2mm}
\begin{center}
\includegraphics[width=80mm]{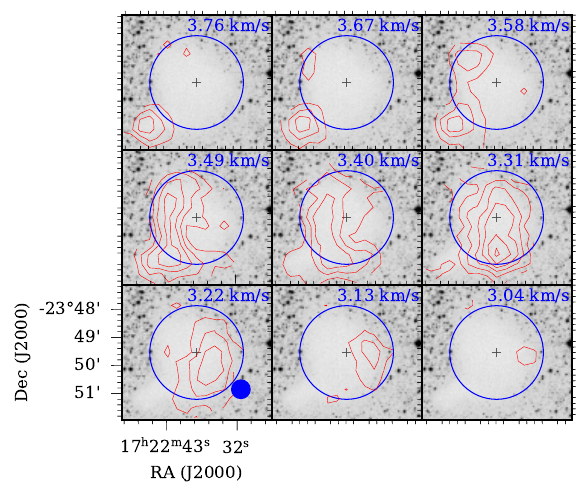}
\end{center}
% \,(1$_{0}$\,-\,0$_{1}$)
\caption{
kinematic component decomposition.
Channel maps of SO (red contours) overlaid on the DSS image (see Fig.\,\ref{spectra_maps}). 
Levels start from 3$\sigma$ ($\sigma$\,=\,0.012\,K\,km\,s$^{-1}$) with an interval of 3$\sigma$.  The
blue circles mark the  main body of B68. The beam size
is shown in the bottom right corner of the bottom left panel.}
\label{chan_maps}
\end{figure}
To investigate the spatial distribution and kinematics of SO, the commonly used moment method was performed to derive 
the velocity-integrated intensity distribution (moment-0), the intensity-weighted velocity field (moment-1) and the intensity-weighted velocity dispersion field (moment-2). The integrated LSR velocities range from 3.0 to 3.8\,km\,s$^{-1}$.
The velocity-integrated intensity (moment-0) maps of B68 are shown in the left panel of Fig.\,\ref{int_maps}. 
The strongest emission of SO is located at the southeastern part instead of the centrally located column density peak detected by the dust extinction measurements with an angular resolution of 10$^{\prime\prime}$ \citep{Alves2001M,Alves2001}. The  central region near the core shows lower velocity integrated intensities. A similar morphology is detected by CS emission line observations \citep{Lada2003}.

The middle panel in Fig.\,\ref{int_maps} shows the velocity field (moment-1) of SO. Three 
velocity components are clearly detected. The western part shows the 
lowest velocity component, and the southeastern part shows the highest velocity component, while the eastern part provides intermediate velocities. The three parts including the bullet and the red- and blueshifed parts of the main body are clearly arising from different regions as can be inferred from the channel maps in Fig.\,\ref{chan_maps}.

The right panel in Fig.\,\ref{int_maps} shows the velocity dispersion (moment-2) of SO. The main core of B68 shows a quite constant velocity dispersion, around 0.09\,km\,s$^{-1}$. The southeastern core shows a larger velocity dispersion, at the order of 0.16\,km\,s$^{-1}$. 

\section{Discussion}
\subsection{Solid-body rotation of B68}
The velocity gradients detected by molecular lines of C$^{18}$O and N$_2$H$^{+}$ in B68 have been explained with 
the solid-body rotation model \citep{Lada2003}. 
The middle panel in Fig.\,\ref{int_maps} shows a similar velocity 
gradient as those seen in C$^{18}$O and N$_2$H$^{+}$ \citep{Lada2003}. Therefore, the solid-body rotation is well traced by the SO. 

The parameters to describe the solid-body rotation can be fitted towards the velocity field measured with SO using the following equation \citep{Goodman1993, Lada2003}
\begin{equation}
     v_{\rm lsr} = v_{\rm sys} +  \frac{\mathrm{d} v}{\mathrm{d} s} \Delta {\rm RAcos(\theta)} + \frac{\mathrm{d} v}{\mathrm{d} s}\Delta {\rm DECsin(\theta)}\,,
\label{eq_rotation}
\end{equation}
where $v_{\rm lsr}$ is the observed line of sight velocity in units of km\,s$^{-1}$, $v_{\rm sys}$ is 
the systemic velocity of the core in units of km\,s$^{-1}$, ${\mathrm{d} v}/{\mathrm{d} s}$ is the velocity gradient in units of km\,s$^{-1}$\,arcsec$^{-1}$, $\theta$ is its direction in units of degrees (e.g., $\theta$\,=\,0 degrees: velocities rise towards the east;
$\theta$\,=\,90 degrees: velocities increase towards the north). $\Delta {\rm RA}$ and $\Delta {\rm DEC}$ are the offsets in right ascension and declination in units of arcsec. 
In summary, $v_{\rm sys}$, ${\mathrm{d} v}/{\mathrm{d} s}$ and $\theta$ 
are the three free parameters and can be determined by fitting the observed velocity field presented in the middle panel of Fig.\,\ref{int_maps}.  

In order to obtain the best-fitting parameters, 
the MCMC (Markov chain Monte Carlo) procedure is used with the python-based code emcee\footnote{https://emcee.readthedocs.io/en/latest/} \citep{Goodman2010}. The implementation details for the emcee code are provided in Appendix B.
The posterior distributions of the three free parameters are shown in Fig.\,\ref{rotation}. The best-fitting parameters are $v_{\rm sys}$\,=3.362\,$\pm$\,0.001\,km\,s$^{-1}$, 
${\mathrm{d} v}/{\mathrm{d} s}$\,=1.79\,$\pm$\,0.02\,m\,s$^{-1}$\,arcsec$^{-1}$ and $\theta$\,=1\hbox{$\,.\!\!^{\circ}$}68\,$\pm$\,0\hbox{$\,.\!\!^{\circ}$}72. 
Compared to the previous results obtained with C$^{18}$O and N$_2$H$^{+}$ \citep{Lada2003}, the velocity gradient  ${\mathrm{d} v}/{\mathrm{d} s}$ and the direction $\theta$  are consistent with the values 1.72\,$\pm$\,0.09\,m\,s$^{-1}$\,arcsec$^{-1}$ for N$_2$H$^{+}$ and 5\hbox{$\,.\!\!^{\circ}$}5\,$\pm$\,1\hbox{$\,.\!\!^{\circ}$}2 for C$^{18}$O, respectively, but there are different with the values 2.25\,$\pm$\,0.024\,m\,s$^{-1}$\,arcsec$^{-1}$ for C$^{18}$O and 23\hbox{$\,.\!\!^{\circ}$}5\,$\pm$\,1\hbox{$\,.\!\!^{\circ}$}2 for N$_2$H$^{+}$. The $v_{\rm sys}$ derived here
is consistent with the values 3.3722\,$\pm$\,0.0018 and 3.3614\,$\pm$\,0.0009\,km\,s$^{-1}$ derived from N$_2$H$^{+}$ and C$^{18}$O, respectively \citep{Lada2003}.

%1\hbox{$\,.\!\!^{\prime\prime}$}65$ \times$1\hbox{$\,.\!\!^{\prime\prime}$}14

\subsection{Radial distribution of  SO abundances}
\begin{figure}[t]
\vspace*{0.2mm}
\begin{center}
\includegraphics[width=90mm]{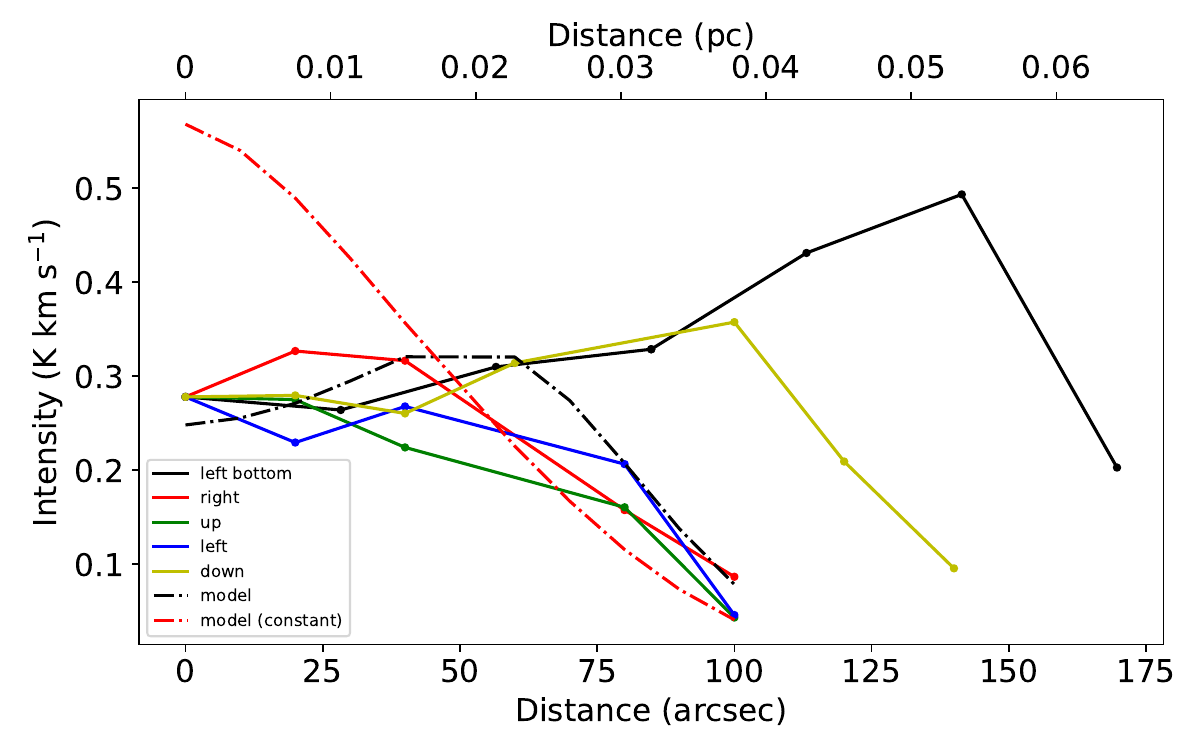}
\end{center}
\caption{Comparison of observed and simulated intensity profiles.
The radial profiles of velocity-integrated intensity from the column density peak (Fig.\,\ref{spectra_maps}) along the five directions shown in the left panel in Fig.\,\ref{int_maps} and the simulated velocity-integrated intensity derived from  the non-LTE radiation transfer code RATRAN \citep{Hogerheijde2000} with non-constant (black dash-dotted line) and constant (red dash-dotted line) SO abundance distributions.}
\label{abundance}
\end{figure}
The abundance profile of SO is derived from the velocity-integrated intensity by performing a non-LTE (Local Thermodynamic Equilibrium) radiation transfer calculation. 
The non-LTE radiation transfer is done with the Monte Carlo code 
RATRAN\footnote{\url{https://sronpersonalpages.nl/~vdtak/ratran/frames.html}}
under 1D spherically symmetry \citep{Hogerheijde2000}. The molecular data file of SO was taken from the 
Leiden Atomic and Molecular Database \citep[LAMDA;][]{Lique2006}
% \footnote{https://home.strw.leidenuniv.nl/$~$moldata/SO.html} 

For the non-LTE radiation transfer calculation, the density structure is  the Bonner-Ebert sphere with  $\xi_{\rm max}\,=\,6.9$ well determined by the dust extinction data \citep{Alves2001}.
% 10\,$\pm$\,1.2\,K 
The gas kinetic temperature is about 10\,K 
derived from NH$_3$ (J,K) = (1,1) and (2,2) data obtained with the Effelsberg 100\,m telescope \citep{Hotzel2002}. Therefore, both the gas and dust temperatures are  set to 10\,K for the calculation. 
From the right panel in Fig.\,\ref{int_maps}, the velocity dispersion for the main core of B68 is about 0.09\,km\,s$^{-1}$.

In Fig.\,\ref{abundance}, the velocity-integrated intensity of SO firstly increases from the core center to then decline towards the boundary. This follows the 
trend of SO velocity-integrated intensity also seen in L1498 and L1517B \citep{Tafalla2006}. 
Therefore, a step function is adopted to describe the radial  abundance profile of SO for B68. SO is depleted at the core center,  while the abundance keeps constant for larger radii \citep{Tafalla2006}.

% The observations of molecular emission lines $^{13}$CO\,(1\,-\,0), C$^{18}$O\,(1\,-\,0) and 
% C$^{18}$O\,(2\,-\,1) have been carried out with the 
% Swedish-ESO Submillimeter Telescope (SEST) \citep{Hotzel2002A}. 

A synthetic SO spectral line cube is produced from the non-LTE radiation transfer. 
Then the cube is convolved with a 40$^{\prime\prime}$ FWHM Gaussian. 
The convolved cube is used to produce velocity-integrated intensities and is then compared to the observational results. Fig.\,\ref{abundance} compares the radial profiles of velocity-integrated intensity with the simulated intensity by RATRAN if the abundances are zero for radii less than 60$^{\prime\prime}$  (0.02\,pc) and 2.0$\times$10$^{-9}$ for radii larger than 60$^{\prime\prime}$, respectively. The depletion radius is about 60$^{\prime\prime}$. Our SO abundance is larger than 
the measured abundances of 4.0$\times$10$^{-10}$ for L1498 and 2.0$\times$10$^{-10}$ for L1517B, respectively.  The optical depth 0.053 is taken  from the 
RATRAN non-LTE simulation. 
In comparison, 
the constant abundance  1.0$\times$10$^{-9}$ extending from the core's center to its boundary was also simulated (see Fig\,.\ref{abundance}). The result revealed that the predicted intensity profile decreases radially from the center to the boundary, showing inconsistency with the observed intensity profile. In conclusion, the step function employed to describe the abundance distribution of SO in B68 demonstrates reliability, and the parameters derived through RATRAN non-LTE simulations prove to be robust.

%Similar with the CS and CO, the SO also is depleted in the core center region. 
%\citep{Tafalla2002,Tafalla2004, Tafalla2006}

% \subsection{Shock waves}

% A set of shock conditions and SO abundance.
% also consider different G0

\subsection{Core-core collision scenario}
\begin{figure}[t]
\vspace*{0.2mm}
\begin{center}
\includegraphics[width=90mm]{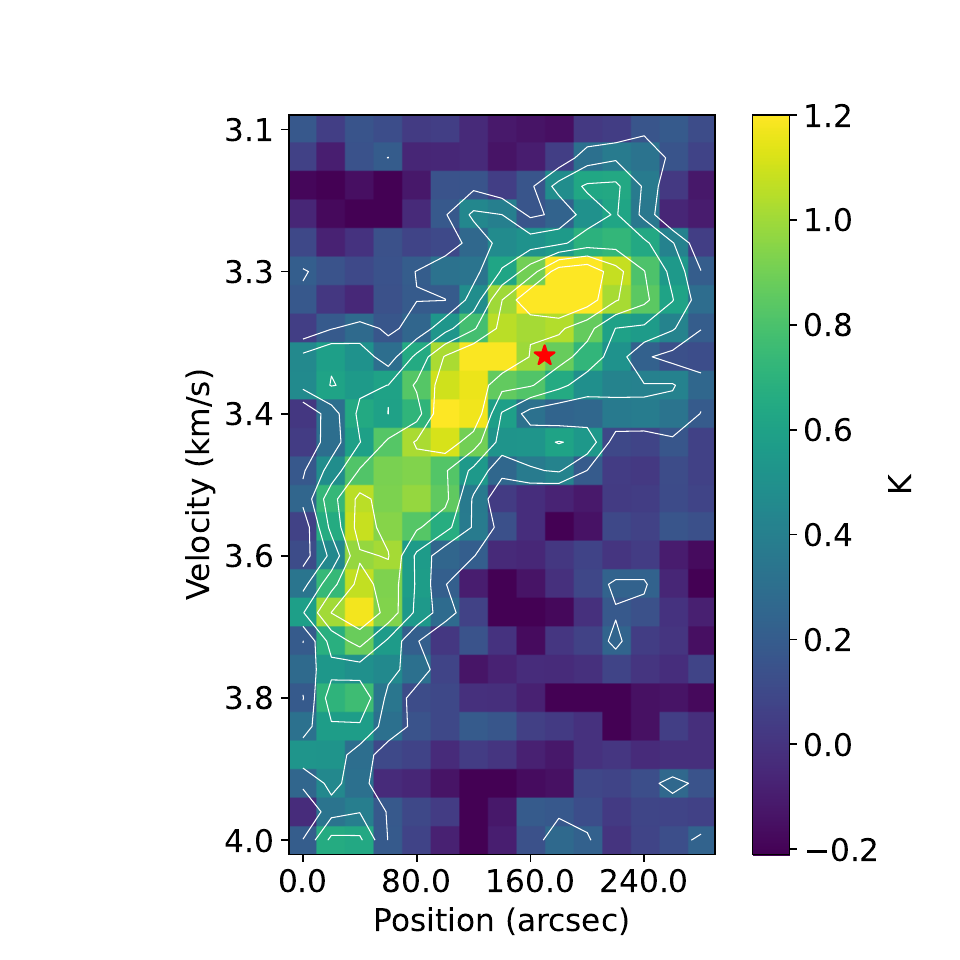}
\end{center}
\caption{Impact velocity estimation between the bullet and the main core.
Position-Velocity diagram along the red straight line from the bottom left  to the upper right shown in the middle panel in Fig.\,\ref{int_maps}. The white contours represent 
the intensity of SO and 
the contour levels start 
from 0.2 to 1.2\,K with an interval of 0.2\,K.
The red star shows the location of the column density peak of the main core.}
\label{pv_diagram}
\end{figure}

From the position-velocity diagram of Fig.\,\ref{pv_diagram}, the velocity difference between the bullet and the main body  is about 0.4\,km\,s$^{-1}$ which is consistent with the simulated collision velocity 0.37\,km\,s$^{-1}$ \citep{Burkert2009}. This provides a conclusive evidence for the bullet impacting on the main source.

In the outskirts of B68, the optical depth of SO is 0.03 estimated by RADEX under conditions of 
$T_{\rm K}$\,=\,10\,K, $n_{\rm H_2}$\,=\,2.0$\times$10$^{4}$\,cm$^{-3}$, 
$N_{\rm SO}$\,=\,1.0$\times$10$^{12}$\,cm$^{-2}$ and line width (FWHM) 0.2\,km\,s$^{-1}$ \citep{vanderTak2007}.
Therefore, SO is  optically thin and the column density is proportional to the velocity-integrated intensity. 
At the position of the bullet the abundance is about five  
times (1.0$\times$10$^{-8}$) that of the abundance at the boundary. The enhanced SO towards the southeastern region (the black line in Fig.\,\ref{abundance}) is likely due to the shocks produced by core-core collision.

In addition, based on the high  spatial resolution of 10$^{\prime\prime}$ dust extinction data, an elongated structure has been detected towards the southeastern part of B68 \citep{Alves2001M,Alves2001}. The elongated structure 
is likely caused by the gravitational tidal force during the core-core collision process  \citep{Burkert2009}. As a matter of fact,  SO is a good tracer to detect this kind of elongated structure caused by gravity \citep{Lee2023}. However, the elongated structure 
is not clearly detected by SO here probably because of the spatial resolution limitation of the data.  

In summary, based on the above kinematical and chemical analyses, the SO observational results 
are consistent with the core-core collision scenario. 

% The line-of-sight infall velocity is 0.12\,km\,s$^{-1}$ from \citet{Lada2003}. 
% If the inclination angle was assumed to be  73$^{\circ}$, the impact velocity is 0.4\,km\,s$^{-1}$.

%  (4.16 - 3.76) 
\section{Summary}
Mapping observations of SO towards the prototypical globule B68 were conducted by the Effelsberg 100\,m telescope with an angular
 resolution of 30$^{\prime\prime}$ (0.01\,pc). 
The results are summarized below:

(1): SO emission is detected both in 
the southeastern bullet and the main body of B68. 
The spectral line profiles are single Gaussian profiles which are quite different when being compared with the complex CS line profiles previously detected.

(2): Based on the velocity-integrated intensity map,  the strongest SO emission appears at the southeastern part of B68 while the central region shows a lower surface density. 

(3): Three velocity components are clearly detected based on the mapped velocity field  and on the SO channel maps. 
The most redshifted is detected towards the southeastern bullet.
Blueshifted emission is detected towards the western part, while emission at intermediate velocities is encountered towards the 
eastern part. 

(4): The measured velocity difference between the bullet and the main core is about 
0.4\,km\,s$^{-1}$ which is comparable with the  velocity 0.37\,km\,s$^{-1}$ predicted by previous numerical simulations. This provides conclusive evidence for the scenario that the bullet is impacting the main core. 

(5): The radial abundance profile is estimated by performing the 
1D spherical non-LTE Monte-Carlo radiative transfer code RATRAN simulating the observed velocity-integrated intensity. 
The abundance stays constant near 2.0$\times$10$^{-9}$ for radii larger than 60$^{\prime\prime}$. The abundance is enhanced 
at the interface between the bullet and the main core. This abundance enhancement is most likely caused by the shock waves produced by the core-core collision.  

In conclusion, based on the kinematic and chemical analyses, 
the observational results suggest that the southeastern 
bullet is impacting on the rotating main core. This is 
consistent with the previously proposed core-core collision 
scenario in B68. 

\acknowledgments
This work was funded by the National Key R\&D Program of China with grant No.2023YFA1608002, by the Youth Innovation Promotion Association CAS, and by the National Natural Science foundation of China (NSFC) under grant nos.12173075 and 12373029. It was also funded by the Chinese Academy of Sciences “Light of West China” Program under grant No. xbzg-zdsys-202212, by the Tianshan Talent Program of Xinjiang Uygur Autonomous Region under grant No. 2022TSYCLJ00055, by the Natural Science Foundation of Xinjiang Uygur Autonomous Region under grant No. 2022D01E06, by the National Key R\&D Program of China under grant No. 2022YFA1603103, by the Tianshan Talent Training Program, by the Regional Collaborative Innovation Project of Xinjiang Uyghur Autonomous Region under grant No. 2022E01050, by the Xinjiang Key Laboratory of Radio Astrophysics under grant No. 2023D04033, by the Chinese Academy of Sciences (CAS) “Light of West China” Program under grant No. 2020-XBQNXZ-017, and by the Science Committee of the Ministry of Science and Higher Education of the Republic of Kazakhstan with grant No. AP13067768.K.W. and T.L. are supported by the Tianchi Talent Program of Xinjiang Uygur autonomous region. C.H. and A. S. have been funded by Chinese Academy of Sciences President's International Fellowship Initiative grants  nos.2025PVA0048 and 2024VMA0002, respectively. 

\bibliographystyle{aasjournal} % style mn2e.bst
\bibliography{ref} % your references Yourfile.bib

\appendix
\section{Bonner-Ebert sphere}
For a spherical ideal gas system with gravity and 
thermal pressure, the system under balance can be described by the Lane-Emden equation. For the isothermal Lane-Emden equation \citep{Chandrasekhar1967}, 
\begin{equation}
\frac{1}{\xi^2}\frac{d}{d\xi}\Bigg(\xi^2\frac{d\psi}{d\xi}\Bigg) = \exp{(-\psi)}\,,
\label{eq1}
 \end{equation}
where $\xi\,=\,(r/C_s)\sqrt{4\pi G\rho_c}$ is the non-dimensional radius, 
$r$ is the dimensional radius, 
$G$ is the gravitational constant, $\rho_c$ is the mass density at
the origin,
$C_s\,=\,\sqrt{k_{\rm B}T/(\mu m_{\rm H})}$ is the sound speed, $k_{\rm B}$ is the 
Boltzmann constant, $T$ is the kinetic gas temperature, $\mu$ is the mean molecular weight of the gas and  $m_{\rm H}$ is the hydrogen mass, and 
$\psi(\xi)\,=\,-\rm ln(\rho/\rho_c)$, $\rho$
is the mass density. The 
boundary conditions of equation (\ref{eq1}) are 
$\psi(0)\,=\,0$ and $d\psi(0)/d\xi\,=\,0$.
Then, the equation (\ref{eq1}) can be numerically solved. 
For a finite radius hydrostatic system, the system can be described 
by a truncated Lane-Emden equation also called Bonnor-Ebert Sphere \citep{Bonnor1956, Ebert1955}. The maximum non-dimensional radius $\xi_{\rm max}\,=\,(R/C_s)\sqrt{4\pi G\rho_c}$, where $R$ is the radius of the system.
\citet{Bonnor1956} and \citet{Ebert1955} studied the 
stability of the truncated Lane-Emden equation. They found the 
system is expected to be gravitational unstable when $\xi_{\rm max}\,>\,6.5$. 
For B68, the density structure can be well fitted by the Lane-Emden equation 
with the $\xi_{\rm max}\,=\,6.9$ indicating that B68 is unstable \citep{Alves2001}.

\section{Procedure for Solid-Body Rotation Fitting}
\begin{figure}[t]
\vspace*{0.2mm}
\begin{center}
\includegraphics[width=90mm]{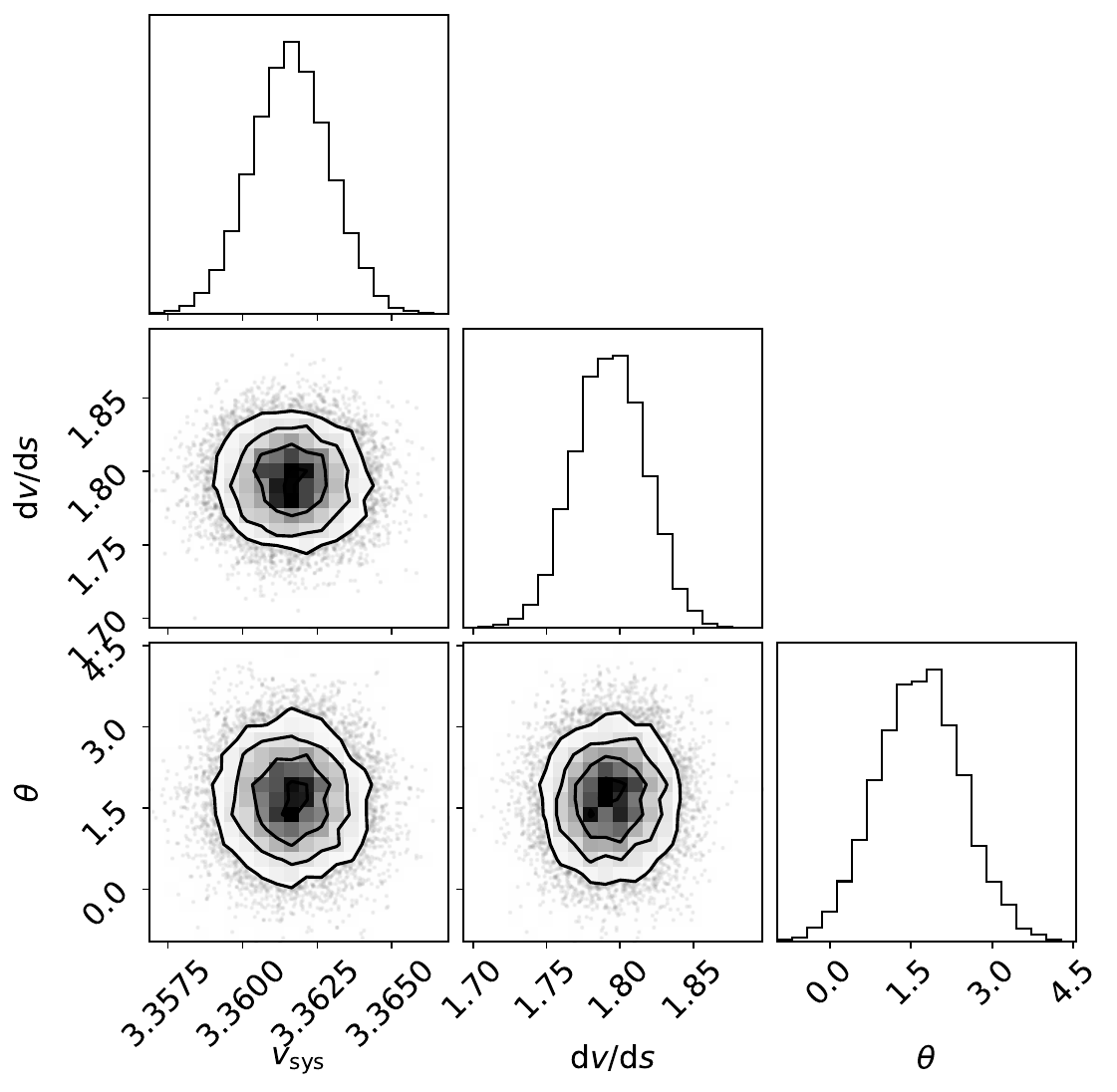}
\end{center}
\caption{The best-fitting results.
The posterior distributions of the three fitted parameters $v_{\rm sys}$, ${\mathrm{d} v}/{\mathrm{d} s}$ in m\,s$^{-1}$\,arcsec$^{-1}$, i.e. the systemic velocity, the velocity gradient, and the direction of the gradient (see eq.\,(\ref{eq_rotation})).}
\label{rotation}
\end{figure}

The emcee code is run by sampling the posterior probability function to conduct the MCMC procedure. The posterior probability function consists of two components: a likelihood function and a prior function. A Gaussian function is adopted as the likelihood function. Therefore, the log likelihood function is defined as  $-0.5\times(v_{\rm obs} - v_{\rm model})^{2}/v_{\rm error}^{2}$, where $v_{\rm obs}$ represents the observed velocity field (i.e., moment-1), $v_{\rm model}$ is the  theoretical velocity field calculated from eq.(\ref{eq_rotation}), and $v_{\rm error}$ denotes the uncertainty of the observed field $v_{\rm obs}$. The typical 
velocity uncertainty $v_{\rm error}$ was calculated as  0.011\,km\,s$^{-1}$ through error propagation analysis 
based on the spectra baseline noise level. 
The prior function is assumed to follow a constant probability distribution. 
The parameter ranges are set as follows: $v_{\rm sys}$  from 3 to 4\,km\,s$^{-1}$, d$v$/d$s$ from 1 to 4\,m\,s$^{-1}$\,arcsec$^{-1}$ and  $\theta$ from -1 to 6$^{^{\circ}}$. 
To run the emcee, 
the parameters nwalkers and steps are set to 32 and 5000, respectively. The calculated autocorrelation time is 
[37, 35, 38]. 
The posterior distributions of the 
three fitted parameters are presented in Fig.\,\ref{rotation}. 

\begin{figure}[t]
\vspace*{0.2mm}
\begin{center}
\includegraphics[width=180mm]{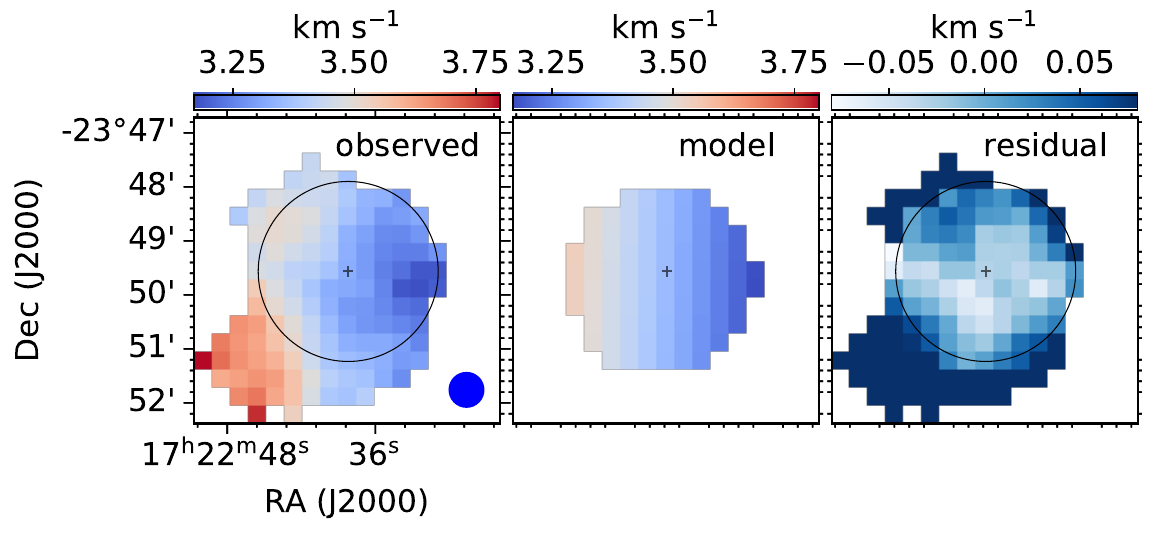}
\end{center}
\caption{Solid-body rotation model assessment. The left panel shows the observed velocity field. The middle panel displays the modeled velocity field derived from the best-fitting parameters. The right panel presents the residual produced by subtracting the model from the observed velocity field. }
\label{residual_rot}
\end{figure}

To evaluate whether the solid-body rotation model is appropriate, Fig.\,\ref{residual_rot} displays the observed velocity field, the modeled velocity field derived from the best-fit parameters, and the residual distribution generated by subtracting the model from the observed velocity field. The residual distribution exhibits remarkable smoothness, with residual values confined to a narrow range of -0.08 to 0.08\,km\,s$^{-1}$. These results demonstrate that the solid-body rotation model is appropriate to describe the velocity field in the main body of B68.

\end{document}